\documentclass[twocolumn]{aastex631}

\usepackage[utf8]{inputenc}
\usepackage{amsmath, amsfonts, amssymb, graphicx}
\usepackage{natbib}
\usepackage{hyperref}
\usepackage{float}

\begin{document}

\title{A Well-Characterized Survey for Centaurs in Pan-STARRS1}
\author[0009-0005-5452-0671]{Jacob A. Kurlander}
\affiliation{Center for Astrophysics — Harvard \& Smithsonian, Cambridge, USA}
\affiliation{DiRAC Institute, Department of Astronomy, University of Washington, 3910 15th Ave NE, Seattle, WA, 98195, USA}
\email{jkurla@uw.edu}
\author[0000-0002-1139-4880]{Matthew J. Holman}
\affiliation{Center for Astrophysics — Harvard \& Smithsonian, Cambridge, USA}

\author[0000-0003-0743-9422]{Pedro H. Bernardinelli}
\affiliation{DiRAC Institute, Department of Astronomy, University of Washington, 3910 15th Ave NE, Seattle, WA, 98195, USA}

\author[0000-0003-1996-9252]{Mario Juri\'{c}}
\affiliation{DiRAC Institute, Department of Astronomy, University of Washington, 3910 15th Ave NE, Seattle, WA, 98195, USA}
\affiliation{Vera C Rubin Observatory, Tucson, AZ}

\author[0000-0003-3313-4921]{Ari Heinze}
\affiliation{DiRAC Institute, Department of Astronomy, University of Washington, 3910 15th Ave NE, Seattle, WA, 98195, USA}
\author[0000-0001-5133-6303]{Matthew J. Payne}
\affiliation{Center for Astrophysics — Harvard \& Smithsonian, Cambridge, USA}

\shorttitle{A Well-Characterized Survey for Centaurs}
\shortauthors{Kurlander et al.}

\begin{abstract}
To prepare for the upcoming Legacy Survey of Space and Time, we develop methods for quantifying the selection function of a wide-field survey as a function of all six orbital parameters and absolute magnitude. We perform a HelioLinC3D search for Centaurs in the Pan-STARRS1 detection catalog and use a synthetic debiasing population to characterize our survey's selection function. We find nine new objects, including Centaur 2010 RJ$_{226}$, among 320 real objects, along with $\sim$70,000 debiasing objects. We use the debiasing population to fit a selection function and apply the selection function to a model Centaur population with literature orbital and size distributions. We confirm the model's marginal distributions but reject its joint distribution, and estimate an intrinsic population of 21,400$^{+3,400}_{-2,800}$ Centaurs with $H_r < 13.7$. The discovery of only nine new objects in archival data verifies that the Pan-STARRS discovery pipeline had high completeness, but also shows that new linking algorithms can contribute even to traditional single-tracklet surveys. As the first systematic application of HelioLinC3D to a survey with extensive sky coverage, this project proves the viability of HelioLinC3D as a discovery algorithm for big-data wide-field surveys. 

\end{abstract}

\section{Introduction}
\subsection{Intrinsic SSO Distributions}
The orbital and physical distribution of Solar System objects (SSOs) is of great significance to both scientific theory and pragmatic interests. The distribution encodes the history of planetary formation and migration processes: the multi-modal orbital distribution of the Kuiper belt provides evidence for and constraints on the migration of Neptune \citep{NeptuneMigration}; depleted areas of the main belt record Jupiter's migration \citep{MainBeltMigration}; and the color distribution as a function of orbit encodes the chemical distribution of the primordial proto-planetary disk, with implications for general planetary formation \citep{CompositionalStructure}. From a planetary defense perspective, the orbital and size distribution of potentially hazardous asteroids determines the frequency of asteroid impacts at different energies \citep{brown2002flux}, with implications for disaster prevention and management \citep{perna2013near}. In each case, the intrinsic distribution of the population is the quantity of interest, rather than the distribution of the sample that has been discovered to date. 

\begin{figure}
    \begin{center}
        \centerline{\includegraphics[width=\linewidth]{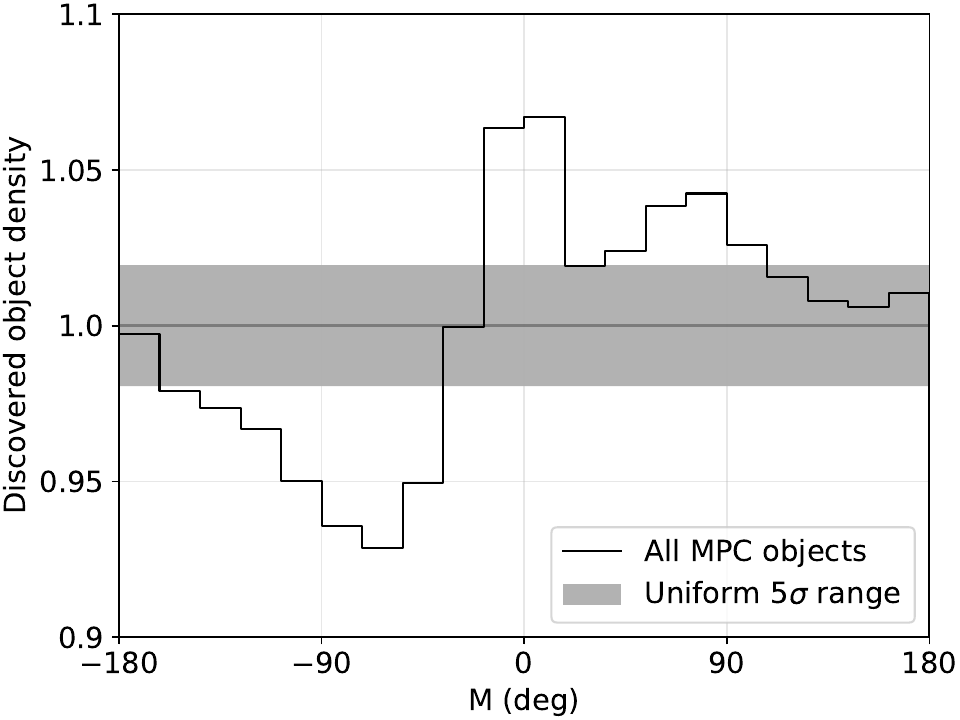} \quad \quad}

        \centerline{\includegraphics[width=1.1 \linewidth]{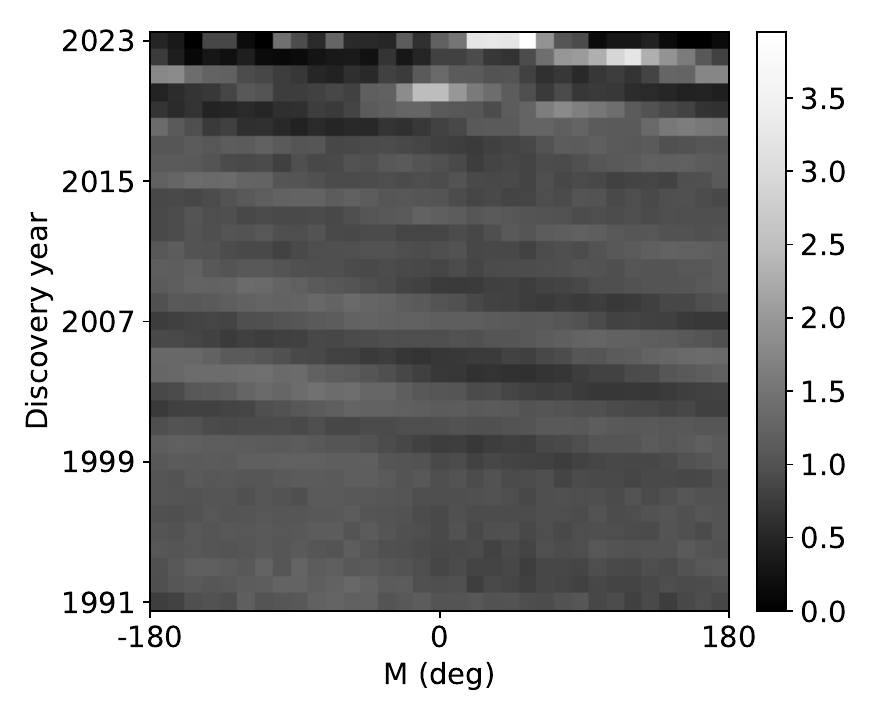}}

        \caption{An unintuitive effect of a simple observational bias on the orbital distribution of discovered objects listed in the Minor Planet Center orbit database.
        \newline
        \textbf{Top panel:} All Minor Planet Center objects by mean anomaly ($M$) as of February 2024. Intuitively, we expect the intrinsic distribution of $M$ in the solar system to be uniform. The peaks and troughs are much greater than the $5\sigma$ ranges of Poisson distributions around the mean, revealing a strong signal explained in the bottom panel. 
        \newline \textbf{Bottom panel:} The cause of $M$ bias is revealed by spreading the data out by discovery date, with color representing the discovery density by mean anomaly within each year (rows have mean 1). At any given time, objects are discovered in a tight region around $M=0$. As time passes, objects increase in $M$, and a year's set of discovered objects broadens as the orbits with different orbital periods drift apart. Orbits of the main belt remain visible as diagonal lines for about two decades, until the drift eventually blurs them together into a uniform distribution with density $1$.}
        
        \label{fig:M_years}
    \end{center}
\end{figure}

However, observational biases skew the distributions of discovered objects away from the intrinsic distribution of objects. Some of these biases are simple: objects which appear brighter are easier to detect, making our completeness on nearby, large, and reflective SSOs far higher than our completeness on distant, small, and dull ones. Figure \ref{fig:M_years} shows an unintuitive bias in the distribution of known SSO orbits which results from a simple observational bias favoring the discovery of objects near pericenter. The observational bias of every SSO search is composed of different non-uniform biases at many steps: observing conditions, a survey's cadence and geometry, mirror and camera effects, and SSO discovery algorithms are each more sensitive to certain orbits than others. The interactions between these biases are not well understood or feasible to approach analytically.

\subsection{Survey Goals}

To prepare to debias the Vera Rubin Observatory's Legacy Survey of Space and Time (LSST), we develop orbit-sensitive debiasing methods. We conduct a HelioLinC \citep{Holman.2018} search in the Pan-STARRS1 (PS1) detection catalog, extract our survey's selection function, and compare our observations to a literature model. 

As a decade-long wide-field SSO survey, PS1 is the available dataset most similar to LSST. As described in \cite{PS1_surveys}, PS1 covered 30,000 square degrees in grizy to a depth of $m_r \sim 21.8$ with four repeat visits per night, though LSST will reach 2.3 magnitudes deeper per exposure, cover its survey area far more frequently, and produce a data volume 60 times larger \citep{LSST_book}. While the discovery of Near-Earth Objects (NEOs) was a primary goal of PS1, it has been a prolific contributor to solar system science in all orbital classes, providing 22\% of all SSO detections between its survey start and February 2024 and discovering the first hyperbolic (interstellar) asteroid: I1/'Oumuamua \citep{Oumuamua}. 

PS1's nearly-uniform cadence and extensive sky coverage makes it an ideal candidate for orbital debiasing, especially for a population with a wide inclination distribution like the Centaurs, which reside between Jupiter and Neptune ($q > 5.2$ au; $a < 30$ au as defined by the Minor Planet Center). \cite{Sisto} note that ``there are few observational estimates of the Centaur population from a well-characterized survey." While Spacewatch \citep{jedicke1997observational} and the Outer Solar System Origins Survey (OSSOS) \citep{OSSOS_VII} were both debiased, both had small survey areas limited close to the ecliptic, and there has not been a debiased Centaur survey with extensive sky coverage to our knowledge. This provides an opportunity to test new debiasing methods while conducting a novel search for Centaurs.

The Centaur population is a subset of a larger, continuous population of objects from the trans-Neptunian region which, through repeated close gravitational encounters with the giant planets, are scattered into more inclined and eccentric orbits than other SSO populations. Although this population is broken up into the scattered disk, the Centaur region, and Jupiter-family comets, these objects originate from the same source \citep{duncan1988origin}, and should share an original size and chemical distribution now modified by the different rates of outgassing at different heliocentric distances. Centaurs have typical mean dynamical lifetimes of only a few Myr \citep{tiscareno2003dynamics}, so they would be depleted quickly if not for a stream of scattered disk objects perturbed into closer orbits \citep{Sisto}. A precise description of the Centaur population would constrain the scattered disk's size distribution for objects too small to directly observe at their distant, present locations. The number and orbits of Centaurs constrain both the scattered disk orbital distribution and the perturbation mechanisms sending scattered disk objects into the Centaur region. 

In this project, we simulate and inject a debiasing population into $\sim$550,000 $r$-band, $w$-band, and $i$-band PS1 exposures ranging from April 2009 to November 2017. We use \texttt{HelioLinC3D}\footnote{See Github: \href{https://github.com/lsst-dm/heliolinc2}{https://github.com/lsst-dm/heliolinc2} \newline Note that we distinguish typographically between the HelioLinC algorithm, the \texttt{HelioLinC3D} implementation of the HelioLinC algorithm, and the \texttt{heliolinc} routine of \texttt{HelioLinC3D.}} (the LSST implementation of the \cite{Holman.2018} HelioLinC algorithm) and the orbit-fitting methods of \cite{Bernstein.2000} with modifications by Bernstein and Bernardinelli\footnote{See Github: \href{https://github.com/gbernstein/orbitspp}{https://github.com/gbernstein/orbitspp}} to find high-quality multi-opposition orbits of real objects simultaneously with synthetic debiasing objects. After using the debiasing objects to fit a survey selection function dependent on apparent magnitude and six orbital parameters, we compare our real discoveries to those ``selected" by our selection function from a model Centaur population and evaluate the model's accuracy. As the first HelioLinC3D search of a wide-field detection catalog, this project also served as an early exploration and stress test. 

The paper is organized as follows. In Section \ref{Debiasing}, we describe the need for orbit-sensitive debiasing methods in the context of previous methods. In Section \ref{Methods}, we detail the generation and injection of our synthetic debiasing population, the search for objects in combined real+synthetic data, and our choice of selection function model. In Section \ref{Results}, we describe the discovered objects, their purity, the extracted selection function and its cross-validation, and compare our discovered objects to a selected literature population. In Section \ref{Discussion}, we discuss our methods and results.

\section{Orbit-Sensitive Debiasing}\label{Debiasing}
\subsection{Survey Debiasing Overview}

The first step in quantifying a survey's bias is to measure its completeness as a function of object parameters (its ``selection function") either by testing a discovery pipeline against synthetic detections injected into exposures or by measuring the detection and non-detection of a catalog of already-known real objects \citep{jedicke1997observational}. Traditionally, selection functions are measured only as a function of apparent magnitude and rate of motion, which determine an object's probability of discovery in single-tracklet surveys and most other discovery mechanisms. In particular, OSSOS \citep{OSSOS_VII} and DES \citep{DES} each debiased catalogs of $\sim 800$ TNOs as a function of apparent magnitude and rate of motion.

The upcoming LSST provides a tremendous opportunity to extend the work done by those deep, well-characterized, but narrow surveys into the wide-field regime. LSST's observations will combine the width of PS1 with the depth of concentrated surveys like DES and OSSOS, and achieve far more repeat coverings than either. LSST will search for SSOs across every orbital class, from NEOs interior to Earth's orbit to TNOs out to at least 150 au, and even interstellar objects (ISOs) not bound to our solar system. It is expected to discover several times more SSOs than are currently known in every dynamical class, vastly expanding our knowledge across the solar system in a single survey with one discovery pipeline.

If LSST's SSO selection function can be well-characterized, its debiased distributions of SSOs would be by far the most broad and precise estimate of intrinsic SSO distributions. These estimates would be the best observational model of the solar system for decades, until some future observatory substantially improves on LSST's breadth and depth.

\subsection{Orbit-Sensitive Debiasing}

One major difference between LSST and previous SSO surveys is its discovery mechanism. While traditional surveys like PS1 discovered objects by submitting single tracklets of three or more detections to the Minor Planet Center, LSST's two-returns-per-night cadence prevents it from producing high-confidence tracklets en masse. Instead, a multi-night linking algorithm, HelioLinC \citep{Holman.2018} will be employed to create high-confidence clusters of three or more low-confidence two-detection tracklets. HelioLinC depends on an assertion of hypothesized orbital parameters (heliocentric distance $r$ and its first two derivatives $\dot{r}$ and $\ddot{r}$), making its behavior inherently less uniform with respect to orbits than traditional methods. For example, it is more effective the closer an object's orbit matches up to an asserted hypothesis orbit. HelioLinC and its interactions with the rest of an SSO survey must be debiased with respect to objects' orbits. 

\begin{figure}
    \begin{center}
        \centerline{\includegraphics[width=\linewidth]{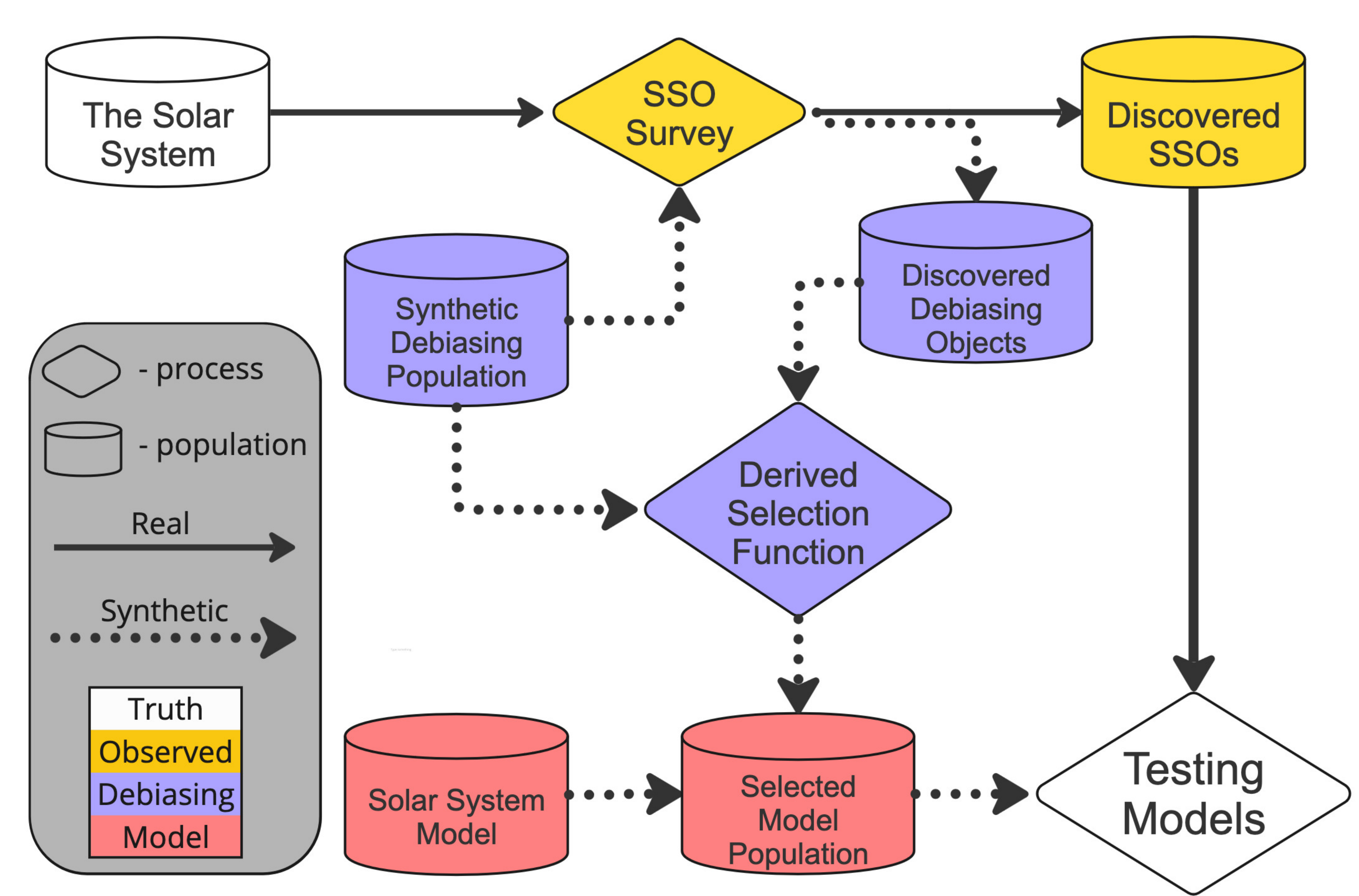}}
    
        \caption{The structure of our SSO survey debiasing method. Using a debiasing population to query the survey's selection function, we can evaluate and reject arbitrarily many model solar systems with only a single survey simulation and discovery process.}

        \label{fig:Flowchart}
    \end{center}
\end{figure}

Unlike single-tracklet discovery methods, running HelioLinC on LSST-scale data is very computationally expensive, making it infeasible to directly apply survey bias by running survey simulations and the real discovery pipeline for many different model solar systems. We also cannot run a single search with a large number of injected populations -- as the density of detections per exposure ($n$) grows, HelioLinC produces a large number of ``mixed" clusters (which do not represent any one object) and degrades the number of pure clusters \citep{Knowlton}. Instead, as shown in Figure \ref{fig:Flowchart}, we inject a single debiasing population to measure the survey's selection function across the entire parameter space of interest. Then, the selection function can easily be applied to many synthetic solar system models, and many selected models can be evaluated against the survey's real discoveries with only one expensive discovery process.

\section{Methods}\label{Methods} 

\subsection{Debiasing Population}

We design a synthetic debiasing population of one million synthetic `debiasing' objects to (a) span the entire orbital range of Centaurs plus some buffer space, (b) have apparent magnitude which span the range of PS1 exposure limiting magnitudes, and (c) be well-distributed throughout the entire parameter space. Object parameters are chosen independently and identically from per-parameter distributions. Pericenter distance ($q$) is chosen uniformly from the interval $(2, 35)$~au, while eccentricity ($e$) is chosen from a normal distribution centered at 0.4 with $\sigma = 0.2$ to match the bulk of discovered Centaurs, and truncated within $(0, 0.99)$. To distribute orbits uniformly over the unit sphere, inclination ($i$) is chosen as the inverse cosine of a uniform distribution over $(-1, 1)$, which distributes objects' angular positions uniformly over the sky. The argument of pericenter ($\omega$), longitude of ascending node ($\Omega$), and mean anomaly ($M$) are each chosen uniformly from $(0, 360)$ degrees. To span the limiting magnitude range of PS1 exposures, an apparent magnitude ($m$) is selected uniformly from $(21, 23.5)$, then transformed into an absolute magnitude ($H$) by asserting that apparent magnitude at the survey midpoint time. 

We assumed an overly-simplified constant Centaur color distribution of $w = r = i$. The effects of this assumption are discussed in Section \ref{color_issue_discussion}. The slope parameter (G) is set uniformly to 0.15 to match Minor Planet Center guidelines. 

We survey-simulate the entire million-object debiasing population, but only consider the $\sim 400,000$ objects in the Centaur region ($q > 5.2$ au; $a < 30.1$ au) when debiasing our Centaur discoveries in Section \ref{Results}. The remaining objects could be used to debias the TNO region out to $q = 35$ au. If this project were conducted again, the debiasing population would extend out to $q \sim 50$ to allow us to debias the entire trans-Neptunian region, eccentricity would be drawn uniformly from $(0, 1)$, and apparent magnitude would extend brighter than 21 to better capture the survey's completeness on bright objects.

The population's motion is simulated from MJD 54466.5 to 58466.5 using a Wisdom-Holman n-body integrator \citep{wisdom1991symplectic} with all planets as perturbers and a timestep of 2 days. Objects with close interactions with perturbers are re-integrated with a timestep of 0.1 days. Then, for fast interpolation and efficient storage, a piecewise Chebyshev polynomial is fit to each object's path, chosen such that the maximum positional error is less than $5 \times 10^{-7}$ au, corresponding to a one-arcsecond error at a distance of 1 au and under a quarter of an arcsecond for objects past Jupiter. Then, observed object positions can be calculated quickly by evaluating the polynomial, along with a light-time correction. 

\subsection{Synthetic Injection and Masking}

To introduce our debiasing population into our survey (see Figure \ref{fig:Flowchart}), we model each object's potential detection in each exposure. We do this in two stages. For each exposure, we first determine which objects in the debiasing population will appear angularly close enough to the center of the exposure to plausibly be detected. For each of these nearby objects, we next examine in detail whether they would have actually been detected by the Pan-STARRS system in that particular exposure. This division cleanly separates the parts that depends only upon the orbits of the debiasing population, the time and direction of the exposure, and the location of the observatory at that time from the parts that depend upon the details of the configuration of the detectors in the focal plane and the characteristics of each detector.

The objects in our debiasing population are sufficiently distant from the Sun and Earth that their apparent motion on the sky is fairly slow, at most a few arcminutes from one night to the next. Thus, it is not necessary to determine the precise locations of all the objects at the times of each exposure. Instead, we calculate the position and corresponding HEALPix\footnote{\href{http://healpix.sf.net}{http://healpix.sf.net}} index (nside=32) of each object at the midpoint of each night. We organize the objects by their HEALPix indices, and for each exposure calculate exact positions only for the $\sim$260 objects in HEALPix which overlap a 1.9-degree circle around the center of the exposure. 

First, we check precisely whether each object falls into the exposure's field of view, then whether it would appear in a cell, subcell, or pixel flagged as `bad' in the PS1 detection pipeline. Then, we fit a t-distributed detection efficiency function to the exposure metadata from the PS1 detection pipeline to evaluate our debiasing sources' probability of $5\sigma$ detection as a function of magnitude. We select objects which fall on live camera area according to their detection probabilities and add astrometric noise based on their signal-to-noise ratios.

Since we debias using only distant ($q > 5.2$ au) objects, we ignore trailing, which can effect magnitude measurements of quickly-moving objects. In the worst-case scenario, an eccentric retrograde Centaur orbit grazing Jupiter's orbit at pericenter could move 0.7 arcseconds in a typical 45-second PS1 image -- half the typical FWHM of 1.3 arcseconds \citep{PSdataprocessing}. In practice, the worst case is never achieved -- the closest discovered debiasing object has a heliocentric distance of 7.2 au, where a worst-case orbit cannot trail more than 0.45 arcseconds. In typical cases, our objects trail around 0.1 arcseconds, less than half the width of a PS1 pixel and safely ignorable for magnitude measurement.

The PS1 detection catalog contains an average of 140,000 detections per exposure -- prohibitively many for \texttt{HelioLinC3D}. This number is dominated by static sources and false positives, so we mask out detections which are very unlikely to represent SSOs. We use a catalog of stationary sources which appear in PS1 five or more times across three or more nights within a half arcsecond radius and mask all detections which lie within 1 arcsecond of a stationary source. Then, we remove the detections which lie in bad subcells and pixels as labeled by the PS1 data pipeline. Finally, we remove all detections from pixels around the border of each chip, where hardware failures at chip edges cause a huge number of false positive detections. With CCD cells defined as $590 \times 598$ grids of pixels, we retain the pixels with $12 \le x \le 581$ and $7 \le y \le 587$. After the full labeling process, detections are distributed fairly uniformly over the cells with some overabundance ($\sim 1.5x$ median) around the edges. Overall, the labeling process removes $92\%$ of the detections and $9.6\%$ of the working area of the camera, down to a more manageable $\sim11,000$ detections per exposure. Synthetic detections are removed through the same process, thereby including the masking process in our debiasing.

\subsection{Object Search}
\begin{table}[h!]\centering
\begin{tabular}{||c c c c||} 
 \hline
 Data product & Total & Real & Debiasing \\ [0.5ex] 
 \hline\hline
 Detections & $6.5 \times 10^9$ & $6.4 \times 10^9$ & $2.4 \times 10^7$ \\ 
 Tracklets & $7.6 \times 10^{10}$ & - & - \\
 Clusters & $7.2 \times 10^8$ & - & - \\
 Linkages & $2.2 \times 10^7$ & - & - \\
 Unique orbit-fit linkages & $7.7 \times 10^5$ & - & - \\
 Pairs & $1.6 \times 10^6$ & - & - \\
 Orbit-fit pairs & $1.9 \times 10^5$ & - & - \\
 Connected components & $7.5 \times 10^{4}$& $329$ & $7.4 \times 10^{4}$\\
 Discoveries & $7.4 \times 10^{4}$ & $320$  &  $7.3 \times 10^{4}$ \\ \hline
\end{tabular}
\caption{Data product volumes}
\label{table:data_volume}
\end{table}

We bin the PS1 detection catalog spatially and temporally (see Section \ref{parallel}) and search each bin in parallel. We construct tracklets of two or more detections, and find plausible clusters of three or more tracklets (see Section \ref{linking}), then perform two orbit-fitting steps to cull our plausible clusters into a set of high-quality linkages (see Sections \ref{linking} and \ref{multi-year}). In order to extend short-arc linkages into multi-year orbits, we pair linkages together in Keplerian coordinates and combine overlapping pairs together (see Section \ref{multi-year}). Volumes of each data product are shown in Table \ref{table:data_volume}. 

\vspace{0.5cm} 

\subsubsection{Parallelization Strategy} \label{parallel}
Even after a 94\% reduction, the number of PS1 detections is too large to run \texttt{HelioLinC3D} on the entire catalog simultaneously. Instead, we must divide the dataset up into time-and-space bins and search each bin in parallel. Because Centaurs have small angular motion over short periods of time, we can ensure that most objects stay within a bin throughout its duration by including a buffer zone around each bin which does not count towards its coverage. We find that a configuration of 684 circles of radius of 0.08 radians can cover the observing area, and with a buffer zone of width 0.12 radians we can feasibly run \texttt{HelioLinC3D} on every bin. The large seasonal gap in observations of any patch of sky makes a convenient temporal cut, so we partition bins by year.

For each bin, we take all exposures from HEALPix which lie within 0.2 radians of its center within the relevant year and concatenate their detections into a single \texttt{HelioLinC3D} input file. The 684 spatial regions and 9 years of data generate 6156 bins, 5634 of which are non-empty. While the regions are equal-area and equal-time, they contain different numbers of exposures and detections, so their processing times in our discovery pipeline vary dramatically.

\subsubsection{HelioLinC3D} \label{linking}

HelioLinC3D is composed of three steps: \texttt{make\_tracklets}, \texttt{heliolinc}, and \texttt{linkrefine}, which we run on each bin in parallel. We run \texttt{make\_tracklets} with a maximum angular velocity of 0.4 degrees per day (much faster than distant objects move across the sky), the default maximum tracklet length of 1.5 hours (twice the typical 45-minute PS1 tracklet duration \citep{MOPS2}), and the default minimum of two detections per tracklet.

Next, we run \texttt{heliolinc}, which propogates tracklets to a single reference time given a hypothesis of $(r, \dot{r}, \ddot{r})$, and exhaustively generates plausible clusters of three or more tracklets within a fixed Cartesian radius. 
Given the high false positive rate among the PS1 detection catalogs, we accept a high false positive clustering rate at this step, as we plan to purify clusters later. 

We search a hypothesis grid composed of the Cartesian product of 44 choices of $r$, 3 choices of $\dot{r}$, and one choice of $\ddot{r}$. To distribute hypotheses uniformly in apparent motion, choices of $r$ are spaced evenly in $r^{-1}$ with 30 hypotheses between $\infty$ and $20$ au, plus an additional 15 hypotheses between $20$ and $10$ au. To cover the entire range of bound orbits, we hypothesize $\dot{r}$ choices of $-1/2$, 0, and $1/2$ times the solar system escape velocity $\sqrt{2GMr^{-1}}$ at each hypothesized $r$. We search this hypothesis grid with two choices of clustering radius (140,000 and 190,000 km) and several reference times to search (spread 14 days apart and collectively covering a bin's timespan) each bin. Recent developments in \texttt{HelioLinC3D}'s cluster purification achieve automatically and more successfully in one search, what we attempted here with the use of multiple parameters, and this portion of the survey would have been conducted differently with access to current code.

The $7.2 \times 10^8$ candidate clusters produced by \texttt{heliolinc} include many real objects, but also have high rates of duplication and false positives. We run \texttt{linkrefine} on each bin to more rigorously vet clusters for plausible orbital fits and remove duplicate and overlapping clusters within bins. We provide \texttt{linkrefine} with a single reference time (the center of each bin's timespan) and clustering radius. \texttt{linkrefine} removes $97\%$ of the \texttt{heliolinc} clusters, leaving us with 22 million higher-quality linkages.

\subsubsection{Multi-Opposition Verification}\label{multi-year}

We perform an additional orbit-fitting step using the routines of \cite{Bernstein.2000} to generate high-precision orbital fits using bin centers as reference angles and times. If a linkage fails, we attempt to fit each of its subsets of $n-1$ tracklets (to a minimum of 3), but do not recurse. After this process, we still have many duplicates or highly-overlapping linkages from separate bins. We remove duplicates manually: after sorting all linkages by fit quality $(\chi^2 / \nu)$, we descend the list, accepting a linkage if fewer than half of its detections are already included in other accepted linkages. After duplicate removal, we are left with 767,499 linkages. 

Finally, we search for sets of two or more orbit-fit linkages which represent the same object by finding nearby pairs of linkages together in Keplerian parameter space, then combining overlapping sets of pairs. In order to match across discontinuities in Keplerian parameter space, we instead consider 9-dimensional $a$, $e$, $i$, $\sin(\omega)$, $\cos(\omega)$, $\sin(\Omega)$, $\cos(\Omega)$, $\sin(M)$, and $\cos (M)$ space, which makes continuous the $360 - 0$ discontinuity in our angular parameters. To account for different parameters' different fit accuracy, we re-scale each dimension by the 95th percentile range of debiasing objects' fits in that dimension. We fit an orbit to the combined data of every pair of linkages within 0.75 in the re-scaled space, again re-trying failures with subsets of $n-1$ tracklets. The resulting successfully-fit pairs of linkages have at minimum 5 tracklets with good orbit-fit residuals ($\chi^2 / \nu < 4 $) with observation arcs of at least 1 year. After using \texttt{NetworkX} \citep{networkx} to combine overlapping sets of pairs into connected components, we are left with a set of high-quality multi-opposition orbits. 

\subsection{Selection Function Fitting}
Next, we use our debiasing population and the population of discovered debiasing objects to fit our survey's selection function (see Figure \ref{fig:Flowchart}). We considered several methods of fitting a selection function to our debiasing population discovery data before deciding to use a nearest-neighbor classifier with objects represented by their Cartesian state vectors and $m$ at survey midpoint time, normalized by each dimension's standard deviation to weight each dimension similarly. In cross-validation, the model better recreates the orbit-dependent selection function than any other method we explored (See Figure \ref{fig:7D_cross_validation}), showing far lower bias. Additionally, the model is quite small and portable, consisting of just one K-D tree and a few arrays which can be easily distributed to researchers looking to test their dynamical simulations against our observations. 

The primary drawback of this model is that it only assigns probabilities of zero or one to different objects, which is unappealing for a few conceptual reasons. Firstly, the model is guaranteed to be poorly calibrated, making its individual predictions untrustworthy. For example, it predicts many events with probability 1 even if those events will not inevitably occur. Second, since the model is deterministic, we cannot take repeated samples from it to generate uncertainty ranges, and instead must draw probabilistic samples from the input population. Third, the selection function is full of discontinuities at every transition between zero and one, and has either zero or undefined gradient everywhere. Finally it is simply bad Bayesian protocol to assign probabilities of zero or one. In practice, the high density of debiasing objects allows this model to perform quite well in cross-validation, so we prefer it for the task of population selection.

\vspace{0.7cm} 

\section{Results}\label{Results}

\subsection{Discovered Objects}

{\begin{table}[h!]
\centering
\begin{tabular}{||c c c c c c c||} 
 \hline
 Object  & a (au) & e & i& H & arc (yr) & \# obs \\ [0.5ex] 
 \hline\hline
 2010 RJ$_{226}$ & 29.8 & 0.58 & 38.8$^\circ$ & 9.9 & 4.4 & 41 \\
 2015 MQ$_{204}$ & 55.2 & 0.35 & 29.7$^\circ$ & 5.6 & 2.3 & 33 \\
 2014 WL$_{616}$ & 46.8 & 0.18 & 24.9$^\circ$ & 5.9 & 2.2 & 23 \\
 2011 WW$_{188}$ & 51.3 & 0.30 & 26.3$^\circ$ & 7.3 & 1.3 & 38 \\ 
 2014 NB$_{96}$ & 41.8 & 0.09 & 12.4$^\circ$ & 6.1 & 3.1 & 20 \\ 
 2014 YD$_{100}$ & 39.9 & 0.30 & 25.3$^\circ$ & 8.0 & 2.9 & 20 \\ 
 2014 HJ$_{344}$ & 39.2 & 0.30 & 18.4$^\circ$ & 7.8 & 2.3 & 22 \\ 
 2014 PZ$_{102}$ & 44.3 & 0.12 & 4.5$^\circ$ & 7.0 & 3.0 & 18 \\ 
 2015 AB$_{309}$ & 39.9 & 0.24 & 24.1$^\circ$ & 7.4 & 1.1 & 21 \\
\hline
\end{tabular}
\caption{Objects newly discovered by this survey. JPL values are given for $a$, $e$, $i$, and $H$ and consider data from other surveys, while observation arc and number of observations are exclusively from this survey. Note the eccentric and highly-inclined Centaur 2010 RJ$_{226}$ and two large TNOs with $H < 6$.}
\label{table:objects}
\end{table}

Our survey yields 74,762 connected components of two or more linkages. We discard the 95 connected components which include either a mix of real and synthetic detections or detections from more than one debiasing object. Of the 74,667 remaining connected components, 74,338 represent a single debiasing object, including 73,215 unique objects, and 329 components which represent real objects, including 320 unique objects. Of these, the Minor Planet Center's MPChecker\footnote{\href{https://minorplanetcenter.net/cgi-bin/checkmp.cgi}{https://minorplanetcenter.net/cgi-bin/checkmp.cgi}} matched 310 closely to known objects at multiple epochs, so we associate them with those real objects. We submitted the 10 which did not associate with known objects to the Minor Planet Center, where nine were newly designated (see Table \ref{table:objects}), while one was associated with the already-known 2011 RS.

The real objects include two Jupiter Trojans, 56 Centaurs and 262 TNOs, including 103 in resonances with Neptune (59 Plutinos), 49 objects scattering off Neptune, 83 classical KBOs, 22 detached. Figure \ref{fig:AEIDiscovered} shows the objects in $a$ vs $e$ and $a$ vs $i$ space, clearly displaying the structure of the Kuiper belt including the 3:2, 2:1, and 5:2 resonant populations. Dynamical classifications within the TNO region are taken from the conveniently-timed \cite{volk2024dynamical}, though five TNOs are still not dynamically classified. Three very distant TNOs were found with $a > 150$, but none with $q > 50$. Two TNOs with extreme inclination are found at $i \sim 70^\circ$ and $110^\circ$. The nine new discoveries include TNOs 2015 MQ$_{204}$ and 2014 WL$_{616}$, which are currently the 221st and 330th largest known SSOs ($H \sim 5.5, 5.9$) and the Centaur 2010 RJ$_{226}$. 2010 RJ$_{226}$ has an extreme orbit for a Centaur -- not only is it the most highly-inclined and third-most-eccentric of the 44 Centaurs we found at $a=29.75,e=0.58,i=38.8^\circ$, but it is has the third-highest semimajor axis of all known Centaurs, is more inclined than 95\% of known Centaurs, and more eccentric than 91\% of known Centaurs. 

Only 44 out of the 56 discovered Centaurs have $m > 21$. The other 12 are brighter than any of our debiasing objects, so we cannot approximate their discovery probability and must exclude them from debiasing. 

\subsection{Purity}
Of the 74,762 connected components, all but 95 are ``pure" -- composed only of detections of a single debiasing object. The 95 ``mixed" linkages include 84 which consist of one object plus a single polluting detection, five which consist of one object plus two or three polluting detections, and six which include many detections each of two debiasing objects which orbit-fit together well. A $6/74762$ rate of mixed linkages corresponds to an expected value of 0.03 mixed linkages among our 320 real discoveries.

\begin{figure*}
    \begin{center}
    
    \centerline{\includegraphics[width=\linewidth]{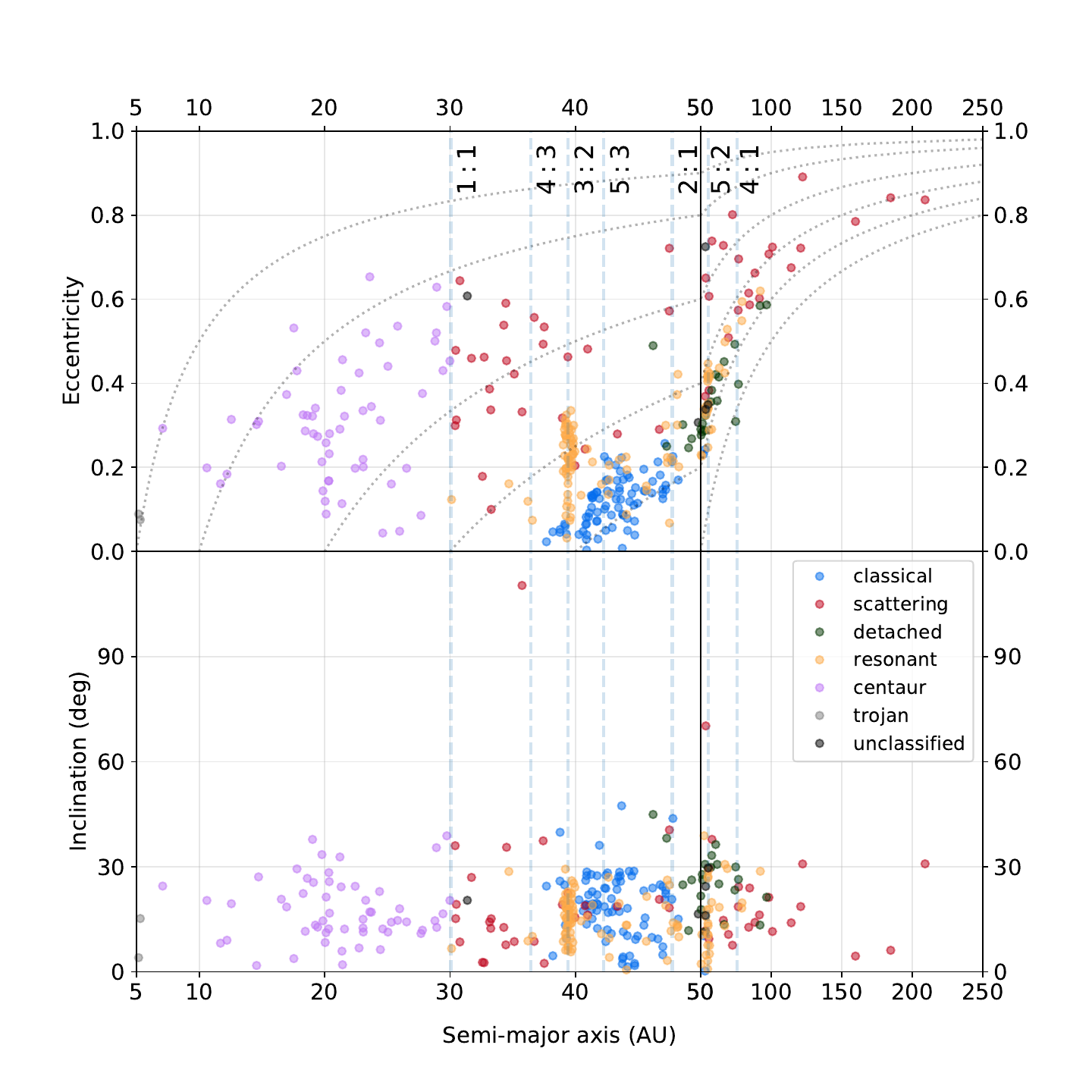}}
    \caption{Real objects discovered by $a$ vs. $e$ (top) and $a$ vs. $i$ (bottom). Resonances of Neptune are shown as vertical dashed lines, while curves of constant perihelion are dotted. Most discovered objects were TNOs rather than Centaurs, though this survey has the largest Centaur yield of any debiased survey to our knowledge.}
    \label{fig:AEIDiscovered}
    \end{center}
\end{figure*}

\subsection{Selection Function}
We find a strong dependence of discovery rate on $m$, $a$ $i$, and $\omega$, but only a moderate dependence on $e$, $\Omega$, and $M$ (see Figure \ref{fig:7D_selection}). Note that these marginal distributions are not independent: we found over 60\% of low-$m$-low-$i$ objects -- far higher than the peak of any marginal distribution. Our selection function approaches zero probability at the dim end allowing us to assume no objects dimmer than $m \sim 23.5$ will be discovered, but since we failed to capture the convergence to a maximum detection probability at the bright end, we can only debias objects with $m > 21$.

We use leave-one-out cross-validation -- predicting each object will behave as its nearest non-self neighbor -- to validate our selection function, predicting which of our debiasing objects were discovered in our search, and comparing those predictions to the set of actually-discovered debiasing objects (see Figure \ref{fig:7D_cross_validation}). Since this method is deterministic, we draw bootstrap samples from our input populations to represent our uncertainty range. Deviations between the true synthetic discovery distribution and the uncertainty range reveal model failures.

\begin{figure*}
    \begin{center}
    
    \centerline{\includegraphics[width=500px]{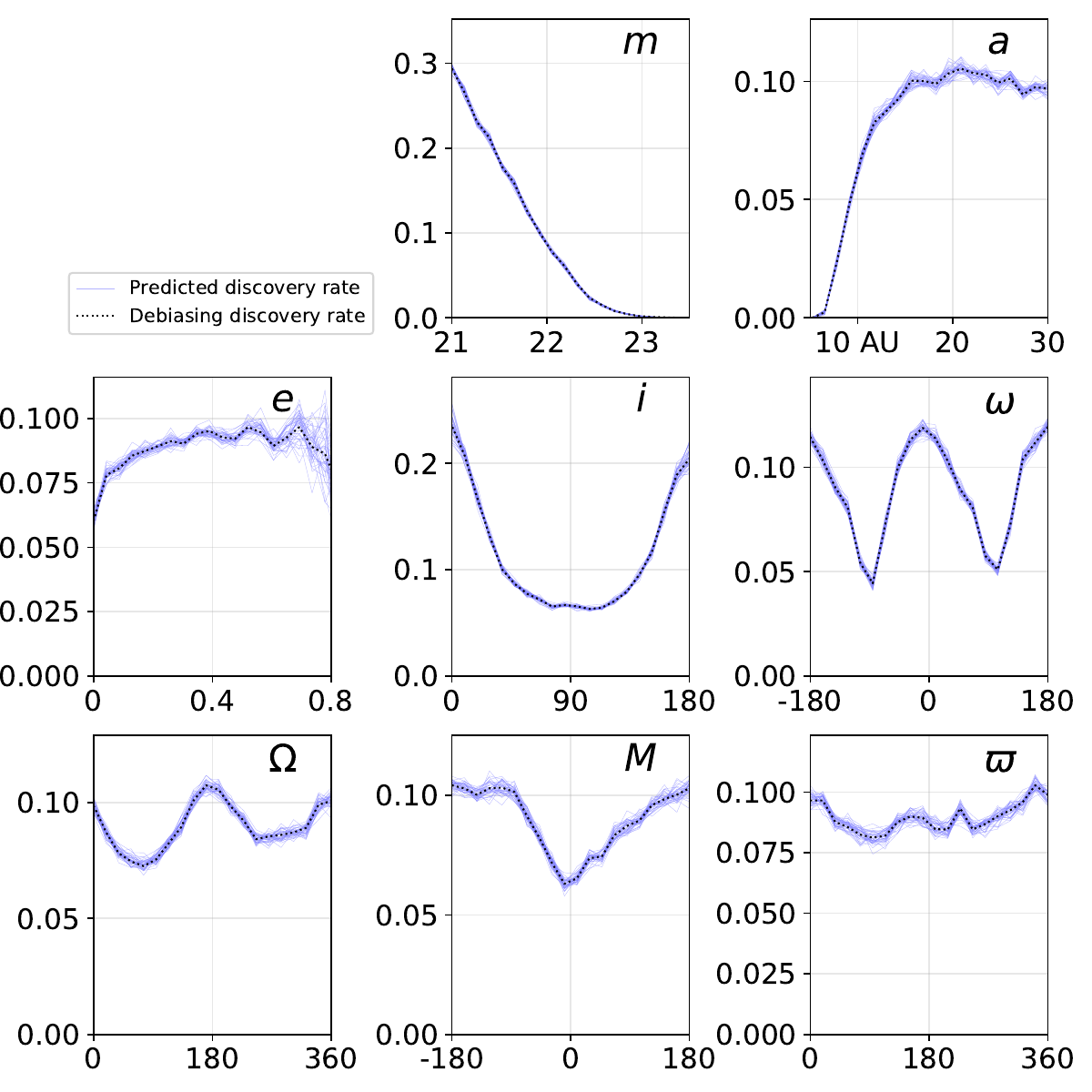}}
    \caption{The selection function marginalized across each dimension. Dotted black lines show the fraction of debiasing objects discovered. Purple lines represent 40 bootstrapped samples of the debiasing population, shown to demonstrate an uncertainty region. Detection probability is strongly dependent on apparent magnitude $(m)$, semi-major axis ($a$), and inclination ($i$), but only moderately dependent on other parameters, and nearly uniform with respect to longitude of perihelion ($\varpi$). Note how the discovery probability at high $m$ converges to zero, but the convergence to a maximum at low $m$ is not captured. The uncertainty region grows where we included few debiasing objects, such as at high $e$ and low and high $i$.}
    
    \label{fig:7D_selection}
    \end{center}
\end{figure*}

\begin{figure*}
    \begin{center}
    
    \centerline{\includegraphics[width=500px]{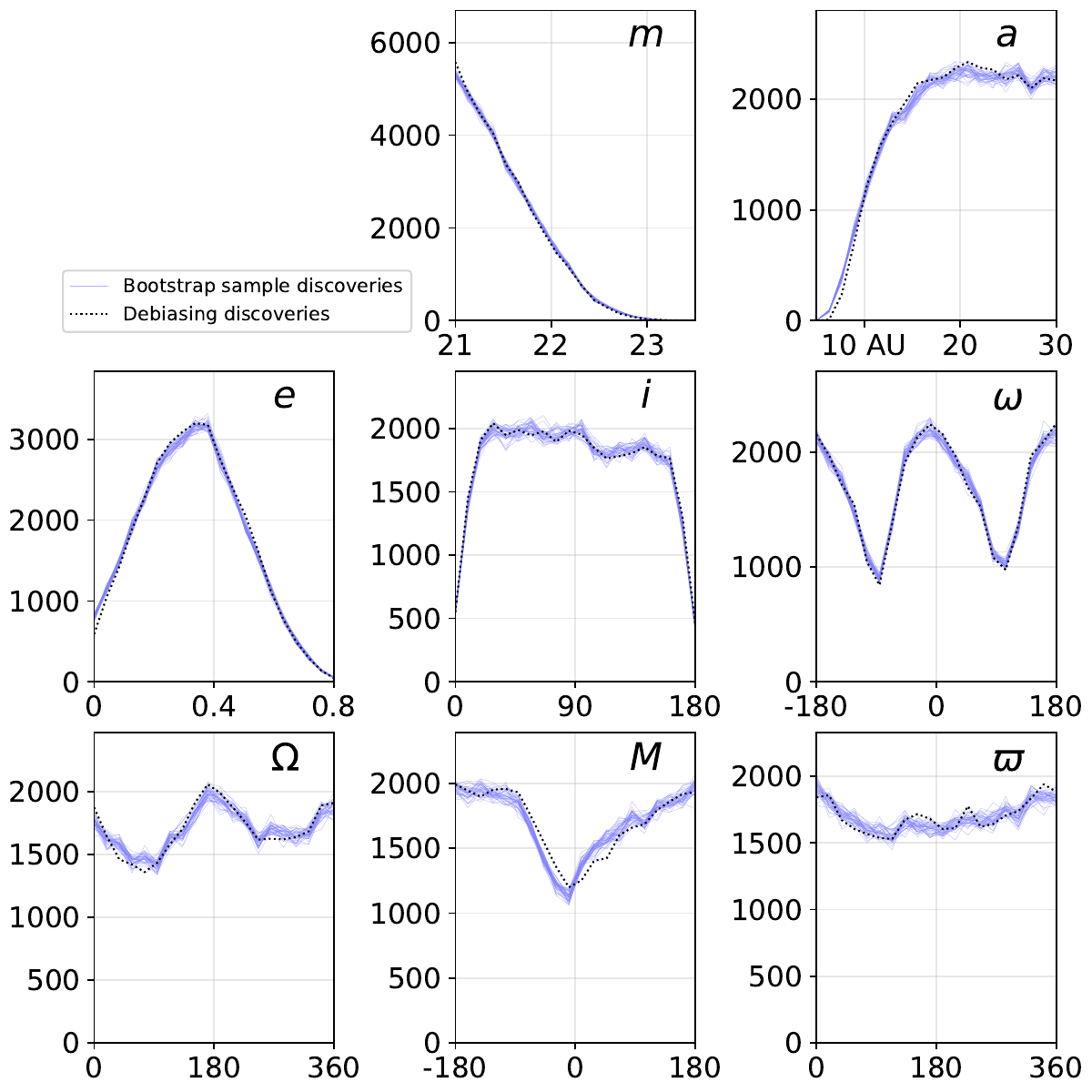}}
    \caption{We perform leave-one-out cross-validation of our model, applying the selection function to 40 bootstrapped samples of the debiasing population. We avoid representing objects as themselves (which would guarantee perfect predictions) by using each object's second-nearest neighbor to predict its detection or non-detection. Overall, the model recreates the marginal distributions well. Deviations between the purple cross-validation region and the true discoveries reveal model bias and shortcomings. In particular, the model is fairly inaccurate in predicting discoveries as a function of $M$. Near the low boundary of $a$ and $e$, the predictions are biased upwards, towards the bulk of the population. Other inconsistent regions include the peaks of $m$ and $a$ and the troughs of $\Omega$.}
    \label{fig:7D_cross_validation}
    \end{center}
\end{figure*}

\vspace{0.8cm} 

\subsection{Literature Model} 
We construct a Centaur population model by combining a literature Centaur dynamical model with a literature Centaur absolute magnitude model, then apply our selection function and compare the resulting selected population to our observations (See Figure \ref{fig:Flowchart}). We use the dynamical model from  \cite{OSSOS_XIX}, which extended work from \cite{NeptuneMigration} to study the scattered disk and Neptune's migration. We adopt the absolute magnitude distribution for the scattered disk and Centaurs from \cite{OSSOS_VIII}, in which a ``knee" power law transitions from a slope of $\alpha_{bright} = 0.9$ to a slope of $\alpha_{faint} = 0.4$ at $H=7.7$, and is restricted to H $ < 13.7$. We apply our selection function to the model Centaurs in the parameter space in which our debiasing objects were distributed (our ``debiasing zone" of $21 < m < 23.5$), and are left with 54,638 (0.209\%) objects which form the selected literature population. 

To estimate the intrinsic number of $H < 13.7$ Centaurs, we model our discovery of debiasing-zone Centaurs as a binomial distribution with $p=0.209\%$ which resulted in 44 real discoveries. This gives a 95\% certainty range of the population size as 15,695 -- 28,227 with 50\% probability at 21,350, leading us to a population estimate of 21,400$^{+3,400}_{-2,800}$. To evaluate the consistency of our observations with the literature model, we perform a KS test between the marginal $a$, $e$, $i$, and $H$ distributions of the selected literature population and our debiasing-zone real discoveries. With p-values of 0.26, 0.39, 0.08, and 0.28 respectively, we find that our data is consistent with the models' marginal distributions at the $p < 0.05 / 4$ level, independently confirming them.

However, we are not limited to comparing the multivariate model and data only with a series of univariate tests. Using the kernel-based KL divergence approximation detailed in appendix \ref{multivariate}, we compute our metric of the typical distances between discovered objects and their nearest model neighbor in a-e-i-H space, and compare the results to the metric on subsamples of the model. We find that the data's divergence is greater than the divergence of 99.64\% of model subsamples, giving a p-value of 0.0036 and indicating a robust disagreement between our data and model. The same metric in a-e-i space gives a non-rejectable p-value of 0.072, indicating that we cannot reject either the dynamical model from \cite{NeptuneMigration} or the size distribution from \cite{OSSOS_VIII} on their own. Figure \ref{fig:heatmap} shows the marginal and 2D joint distributions of our discovered debiasing objects and real discoveries.

\begin{figure*}
    \centering
    \includegraphics[width=\linewidth]{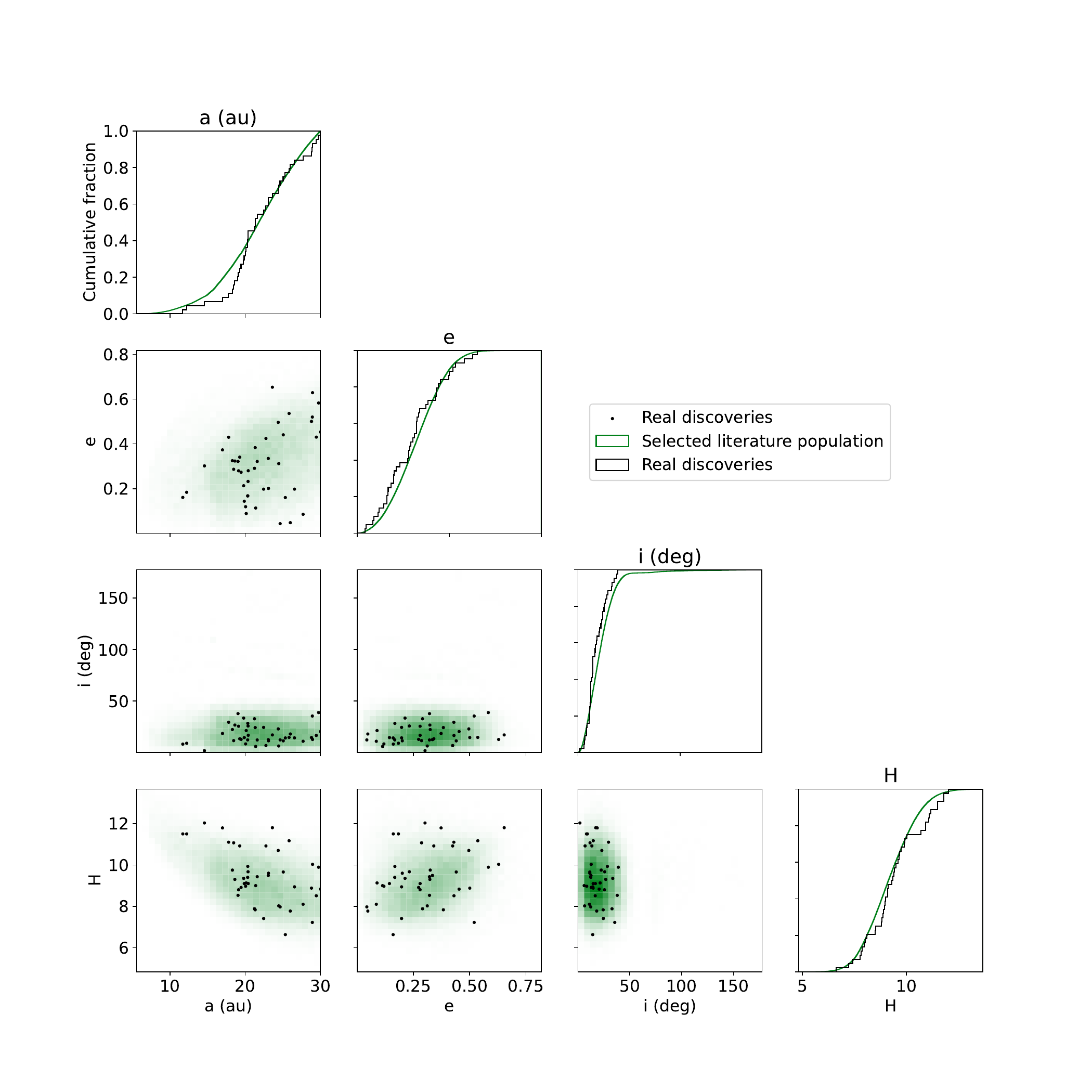}
    \caption{Discovered objects (black) scattered over a heatmap of the selected literature population in each pair of \{a, e, i, H\}. Plots on the diagonal show the mariginal CDFs across each dimension. Visually, the populations appear to be in agreement, though the dearth of discovered objects around $a \sim 17$, $H \sim 10.5$, and of objects with both low $a$ and low $H$ are each noteworthy.}
    \label{fig:heatmap}
\end{figure*}

\vspace{0.8cm} 

\section{Discussion}\label{Discussion}
\subsection{HelioLinC Search}
As the first systematic application of HelioLinC to a wide-field survey\footnote{\cite{Holman.2018} used HelioLinC on the Minor Planet Center's high-quality Isolated Tracklet File while \cite{Yasin} was performed on a small-area, high-cadence survey with DECam \citep{DECamDDF}.}, this project proves the viability of HelioLinC as a discovery tool for wide-field SSO surveys. The discovery of nine objects which were not already found by the PS1 MOPS system demonstrates that HelioLinC can discover more objects than traditional discovery pipelines even in noisy survey environments. Substantial improvements to \texttt{HelioLinC3D} have been made since our search, and it is still under active development and optimization. 

The most obvious drawback of our discovery method is that we make many stringent cuts in series to purify our $\sim 10^9$ HelioLinC clusters into high-confidence multi-opposition orbits. Even with high completeness at each linking and purification step, a large fraction of objects might have been lost. It is very likely that we found and discarded new single-opposition orbits due to our large volume of spurious HelioLinC clusters, which was due to the large volume of spurious PS1 detections. Future HelioLinC surveys can avoid this issue by including fewer and more robust refinement steps.

The small number of new objects compared to the number of re-discovered objects is expected: since PS1 data was already processed for moving objects, we expected that most objects which formed tracklets would have been already discovered by MOPS, if not by \cite{Holman.2018}. We verify that PS1's discovery pipeline had high completeness, but the discovery of nine new objects shows that new linking algorithms can contribute to traditional single-tracklet surveys.

\subsection{Fast Selection Function Model}
The nearest-neighbors selection function appears to be quite effective, encoding a good proxy for our survey's selection function with fast runtime ($>$ 20,000 objects per second) and small filesize ($<60$MB). The model, along with an example notebook introducing its use, is available here, allowing readers to test their own dynamical Centaur or TNO models against our observations: \href{https://doi.org/10.5281/zenodo.14201491}{https://doi.org/10.5281/zenodo.14201491}. To the best of our knowledge, there has been no other published debiasing of outer solar system discoveries from PS1, though \cite{weryk2016distant} suggests it as a future work and the project discussed by \cite{lilly2017debiasing} is ongoing.

In cross-validation (Figure \ref{fig:7D_cross_validation}), our model generates populations which are consistently smaller or larger than the true population in some parameter regions. This is expected in regimes where objects are queried by objects dissimilar to them, such as in very low-density regimes, near parameter space boundaries, where the density of debiasing objects is non-uniform, and at local extrema. In particular, since we include very few objects at low or high eccentricity and inclination, or at low semi-major axis, these regions should be poorly represented in our model, and their neighbors will preferentially lie towards the peak of the distribution. In fact, we do see these issues. For example, among the 708 extreme debiasing Centaurs with $a < 6$~au, 97\% have a nearest neighbor with greater $a$, creating a strong bias where nearby objects are proxied by more distant ones. By comparison, debiasing objects among the 17,744 with $20$~au~$< a < $~21~au have a 49\% chance of being proxied by a more distant object, creating no strong bias. Additionally, there is a major discrepancy between the cross-validation and true distributions of discovered debiasing objects by $M$, indicating that objects are proxied by dissimilar objects, though we marginalize over this parameter for science purposes in this paper. Improving these issues is a high priority for future work on this topic, but we believe the error in $M$ and few-percent errors on other parameters are acceptable for this experiment given its small sample size of real objects and resulting wide statistical bounds. 

Some of the model's cross-validation error could be mitigated by a better choice of debiasing population and parameter space for nearest-neighbor matching. In the future, debiasing populations should be distributed uniformly in the selection function's parameter space so that the selection function is queried on average at the location of an input object rather than preferentially towards the mass of the debiasing population. We will further investigate the best choice of selection function parameter space and debiasing population in future works.

\subsection{Intrinsic Centaur Population Statistics} \label{color_issue_discussion}
Our population estimate of 21,400$^{+3,400}_{-2,800}$ Centaurs with $H_r < 13.7$ is consistent with the OSSOS estimate of 21,000 $\pm$ 8,000 \citep{OSSOS_XIX}. Our discovered orbital and $H$ distributions are consistent with the \cite{OSSOS_XIX} dynamical model and \cite{OSSOS_VIII} size distribution individually, but not with their joint distribution. Inspection of Figure \ref{fig:heatmap} shows a few discrepancies between our detected population and the model -- we observe fewer Centaurs than the model expects with $a < 18$ au, and a higher density with $18$~au~$ < a < 20.5$~au. We also observe more Centaurs than the model expects with $10$~deg~$ < i < $~15~deg. However, these discrepancies are not statistically powerful enough to reject the model distributions.

The eventual purpose of the debiasing tools presented here is to iterate through different choices of input model and eventually produce a model consistent with the data, though doing so here is beyond the scope of this paper. For example, testing alternative TNO size distributions for consistency with our data could provide insights about the link between the TNO and Centaur, but we leave these insights for future debiased surveys with larger sample sizes.

Our debiasing of the $H_r$ distribution (and therefore our estimate of the number with $H_r < 13.7$) is dependent on our Centaur color choice of $r=w=i$. To test this choice, we fit our own $H_r$, $H_w$ and $H_i$ to all Centaurs with PS1 observations in $r$, $w$, and $i$. We find typical colors of $r = i + 0.24 \pm 0.19$ and $r = w - 0.07 \pm 0.14$ -- not significantly different from $r=w=i$. Since observations in each band contributed to our discovery and some objects were found with detections in a combination of filters, there is no simple offset we can make to our $H_r$ distribution to rectify the discrepancy, so we do not adjust our population estimate. However, one possible correction would be to offset our $r$-band magnitudes to the average of $r$, $w$, and $i$, which corresponds to a change of of 0.05 magnitudes and a $5\%$, or 1,000-object reduction in the population estimate. 

\subsection{Application of these Methods to Other Surveys}
The orbit-sensitive debiasing methods used for this survey are fairly general, and are appropriate for any SSO survey using a linking algorithm where an object's orbital parameters beyond apparent rate of motion and magnitude affect its probability of discovery. Our multivariate hypothesis test is general and can compare any observed and model populations. We encourage the use of multivariate tests rather than multiple marginal tests whenever parameters have meaningful joint distributions.

We plan to use a similar process to debias LSST. The debiasing zone for LSST must include NEOs interior to Earth's orbit, extreme TNOs as distant as tracklets can be generated ($\sim$ 150 au), and even ISOs. We plan to debias five colors and light curve amplitudes and periods in addition to orbits and sizes. How to balance the coverage, size, and uniformity of a 14-dimensional debiasing population throughout the entire solar system without overwhelming the discovery pipeline is an open problem. If we can do it appropriately, the debiased LSST catalog will provide unprecedented constraints on intrinsic distributions. Using DP0.3\footnote{Access via the Rubin Science Platform: \href{https://data.lsst.cloud}{https://data.lsst.cloud}.} numbers, we expect LSST to discover $\sim 2 \times 10^4$ TNOs and $4 \times 10^6$ main belt asteroids. Where this project's Poisson uncertainty range allowed for $\sim 15\%$ predictions about Centaurs, a debiased LSST yield could make predictions about TNOs with $0.6\%$ uncertainty, main belt asteroids with 0.05\% uncertainty, or $0.5\%$ uncertainty on 100 different main belt subpopulations. Models of the main belt could be rejected at $5\sigma$ based on percent-level discrepancies on percent-level tails. The asteroid impact rate as a function of absolute magnitude could be determined very precisely, and new precision solar system science would be enabled.


\appendix
\section{Multivariate Hypothesis Tests} \label{multivariate}

\begin{figure*}
    \centering

    \includegraphics[width=0.8 \linewidth]{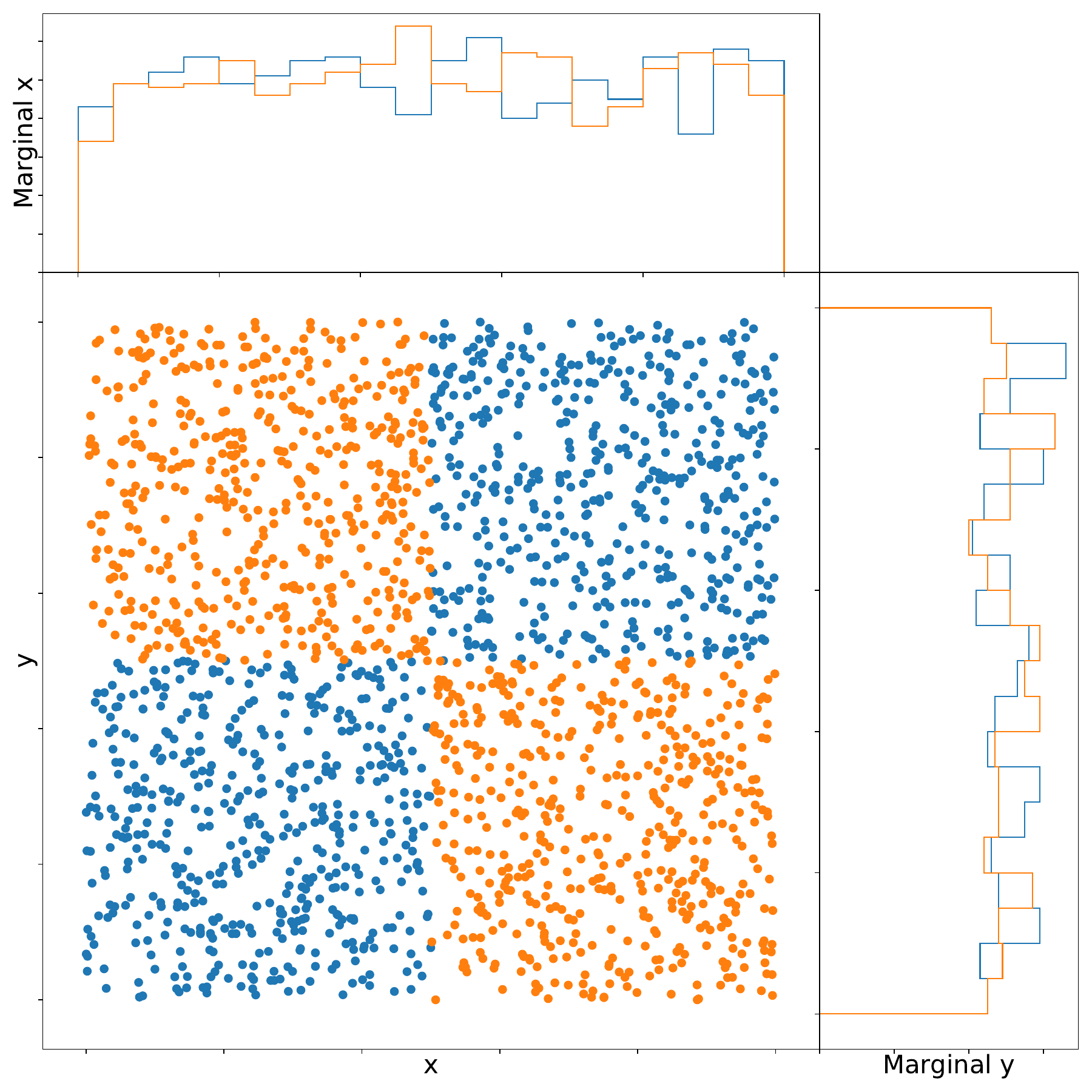}

    \caption{A toy example showing realizations of two distinct distributions which share identical marginal distributions in two quantities of interest (``$x$'' and ``$y$''), but have distinct two-dimensional distributions. A one-dimensional test such as the KS test would fail to distinguish between these two populations, motivating a need for multivariate statistics.}
    \label{fig:joint_marginal}
\end{figure*}

In outer solar system science, the typical comparison between a catalog $\mathcal{D} = \{a, e, i, \Omega, \omega, M, H\}$ sourced from a population $p(a,e,i,\Omega,\omega, M,H)$ and a model $q(a,e,i,\Omega,\omega, M,H)$ is performed with tests such as the Kolmogorov-Smirnov test (or its variants) on the marginal distribution of each parameter of interest. Unfortunately, such tests are not sensitive to discrepancies which are present in the joint distribution of features, but not in any one marginal distribution (see Figure \ref{fig:joint_marginal} for a toy example). We use a divergence-based two-sample test to discriminate between joint distributions. As is common in machine learning literature, we use the Kullback-Leibler (KL) divergence \citep{KL}. For a $n$-dimensional quantity $\mathbf{x}$, we have that the divergence between distributions $p$ and $q$ is 
\begin{equation}
     D_\mathrm{KL}(p||q) \equiv \int_{\mathbb{R}^n} \mathrm{d}\mathbf{x} \, p(\mathbf{x}) \log\left(\frac{p(\mathbf{x})}{q(\mathbf{x})}\right).
     \end{equation}
This divergence is 0 if $p$ and $q$ are identical, and increasing values of $D_{\mathrm{KL}}$ correspond to increasing `distances' between $p$ and $q$.

\cite{nearestneighbor1} and \cite{nearestneighbor2} proposed a kernel-based approximation of this quantity. For two datasets in $d$ dimensions $\{\mathbf{x}_1, \dots,\mathbf{x}_n\}$ sourced from distribution $p$ and $\{\mathbf{x}'_1, \dots,\mathbf{x}_m'\}$ sourced from $q$, we have that 
\begin{equation}
    \hat{D}_\mathrm{KL}(p||q) = \frac{d}{n} \sum_i \log \frac{r_k(\mathbf{x}_i)}{s_{k}(\mathbf{x}_i)} + \log\frac{m}{n-1},
\end{equation}
where
\begin{equation} r_k(\mathbf{x}_i) = \min^{(k)}_{j} || \mathbf{x}_i - \mathbf{x}_j ||
\end{equation}
and 
\begin{equation}
    s_k(\mathbf{x}_i) = \min^{(k)}_{j} || \mathbf{x}_i - \mathbf{x}'_j || 
\end{equation}
where $\min^{(k)}$ is the minimum distance to the $k^{\text{th}}$ nearest neighbor of $\mathbf{x}_i$. These quantities can be readily evaluated using k-d trees, so this is a computationally inexpensive test. 

Using this metric we perform a hypothesis test: the value of $\hat{D}_\mathrm{KL}$ by itself is hard to interpret, but we can derive an empirical $p$-value using the standard Monte Carlo method of repeatedly drawing $n$ samples from the set $\{\mathbf{x}'\}$, and computing the $\hat{D}'_\mathrm{KL}$. The $p$-value, then, is the frequency with which $\hat{D}_\mathrm{KL} \leq \hat{D}'_\mathrm{KL}$, and can be used and interpreted the same way as a $p$-value from a univariate hypothesis test.

In the linked repository, we include a notebook with a more detailed derivation, an implementation of the metric, a demonstration that the metrics have uniformly-distributed p-values on data drawn from a model, and the data and code which generated the p-values and plots: \href{https://doi.org/10.5281/zenodo.14201491}{https://doi.org/10.5281/zenodo.14201491} or \href{https://github.com/Gerenjie/Survey-Debiasing}{https://github.com/Gerenjie/Survey-Debiasing}.

\section*{Acknowledgements}

J.K. acknowledges the support from the University of Washington College of Arts and Sciences Department of Astronomy and thanks the LSST-DA Data Science Fellowship Program, which is funded by LSST-DA, the Brinson Foundation, and the Moore Foundation; his participation in the program has benefited this work.

J.K., P.H.B., and M.J. acknowledge support from the DIRAC Institute in the Department of Astronomy at the University of Washington. The DIRAC Institute is supported through generous gifts from the Charles and Lisa Simonyi Fund for Arts and Sciences, and the Washington Research Foundation.

MJP acknowledges support from the NASA Minor Planet Center Award 80NSSC22M0024 and from NASA award 80NSSC20K0641. 

This work made use of the following software packages: \texttt{astropy} \citep{astropy:2013, astropy:2018, astropy:2022}, \texttt{Jupyter} \citep{2007CSE.....9c..21P, kluyver2016jupyter}, \texttt{matplotlib} \citep{Hunter:2007}, \texttt{numpy} \citep{numpy}, \texttt{pandas} \citep{mckinney-proc-scipy-2010, pandas_6702671}, \texttt{python} \citep{python}, and \texttt{scipy} \citep{2020SciPy-NMeth, scipy_5725464}.
Some of the results in this paper have been derived using \texttt{healpy} and the HEALPix package\footnote{http://healpix.sourceforge.net} \citep{Zonca2019, 2005ApJ...622..759G, healpy_5012376}.

Software citation information aggregated using \texttt{\href{https://www.tomwagg.com/software-citation-station/}{The Software Citation Station}} \citep{software-citation-station-paper, software-citation-station-zenodo}.

The Pan-STARRS 1 Surveys (PS1) have been made possible through contributions of the Institute for Astronomy, the University of Hawaii, the Pan-STARRS Project Office, the Max-Planck Society and its participating institutes, the Max Planck Institute for Astronomy, Heidelberg and the Max Planck Institute for Extraterrestrial Physics, Garching, The Johns Hopkins University, Durham University, the University of Edinburgh, Queen’s University Belfast, the Harvard-Smithsonian Center for Astrophysics, the Las Cumbres Observatory Global Telescope Network Incorporated, the National Central University of Taiwan, the Space Telescope Science Institute, the National Aeronautics and Space Administration under grant No. NNX08AR22G issued through the Planetary Science Division of the NASA Science Mission Directorate, the National Science Foundation under grant No. AST-1238877, the University of Maryland, and Eotvos Lorand University (ELTE).

\bibliography{refs}{}

\end{document}